\documentclass[a4paper]{jpconf}
\usepackage{graphicx}
\usepackage{amsmath,amsthm,latexsym,amssymb,amsfonts,epsfig, psfrag,url,color}
\usepackage{dsfont}
\def\be{\begin{equation}}
\def\ee{\end{equation}}
\def\ba{\begin{eqnarray}}
\def\ea{\end{eqnarray}}

\newcommand{\cp}{\mathcal P}
\newcommand{\cq}{\mathcal Q}

\begin{document}
\title{Canonical formalism for simplicial gravity\footnote{Based on a talk by the author at the conference {\it Loops '11} in Madrid on May 24th 2011 \cite{talk}.}}

\author{Philipp A H\"ohn}

\address{Institute for Theoretical Physics, Universiteit Utrecht, Leuvenlaan 4, NL-3584 CE Utrecht, The Netherlands}

\ead{p.a.hohn@uu.nl}

\begin{abstract}
We summarise a recently introduced general canonical formulation of discrete systems which is fully equivalent to the covariant formalism. This framework can handle varying phase space dimensions and is applied to simplicial gravity in particular.
\end{abstract}

\section{Motivation and Goal}

Research in simplicial gravity\footnote{We refer to discrete gravity theories based on simplices and triangulations.} has mainly been carried out in the covariant setting which directly involves an action principle, while efforts to construct a suitable canonical formalism have been few and far between. For Regge Calculus \cite{regge}, a specific discretization of General Relativity, first attempts to construct a canonical formulation \cite{piwill86} relied on discretized space, but a continuous (infinitesimal) time evolution generated by a set of constraints. A consistent canonical formulation, however, should reproduce exactly the covariant solutions following from the action, which in simplicial gravity implies a discrete time evolution that can thus no longer be generated by a set of constraints employing a bracket structure. Rather, a well--defined set of evolution moves is required to generate such a discrete time evolution. The role of the constraints (if existent) in such a framework, in turn, is reduced to generating (infinitesimal) transformations of the phase space variables which under certain conditions correspond to gauge symmetries of the (discrete) action. Based on ideas put forward in \cite{gambpul03,tent,bd1}, the first canonical formulation (for Regge triangulations) which reproduces exactly the covariant theory and encompasses a discrete version of a hypersurface deformation algebra appeared in \cite{dithoe1}. The drawback of this first canonical formulation is, however, that the evolution moves employed in the discrete time evolution belong to a special class \cite{tent} that preserve both the topology and connectivity of the `spatial'\footnote{For simplicity, we work in Euclidean signature. Hence, we write `spatial' in quotation marks.} triangulated hypersurfaces which carry the phase space variables. This formalism is therefore only applicable to a very special class of triangulations with preserved phase space dimension. Instead, in order to deal with arbitrary triangulations (of fixed `spatial' topolgy), a general class of evolution moves as well as a new canonical framework is necessary.

Such a general canonical formalism for discrete theories which can handle varying phase space dimensions and constraints and, furthermore, is fully equivalent to the covariant formulation has been introduced recently \cite{csg}. In particular, it has been applied to (Euclidean) Regge Calculus with arbitrary `spatial' hypersurfaces.

Here we want to summarise how one may treat the general situation where the `spatial' lattice evolves/changes (akin to the situation with graphs in Loop Quantum Gravity) \cite{csg}. Specifically, this implies that the numbers of degrees of freedom may vary from step to step, a feature which we must be able to incorporate into our formalism. For several physical situations (e.g.\ expanding or contracting universes) and their numerical implementation it may be advantageous to be able to adapt the discretization density in time. But also in the light of several approaches to quantum gravity---especially Loop Quantum Gravity which entails changing spatial graphs and the simplicial Spin Foam models---it seems fruitful to be able to handle evolving lattices.

\section{Evolution in discrete `multi-fingered' or `bubble' time}

\begin{figure}[htbp!]
\begin{minipage}{1.6in}\begin{center}             
   \psfrag{S}{\footnotesize $\Sigma_k$}
   \psfrag{Sf}{\footnotesize$S_{k-}$}
   \psfrag{Sp}{\footnotesize\textcolor{white}{$S_{k+}$}}
   \includegraphics[scale=.19]{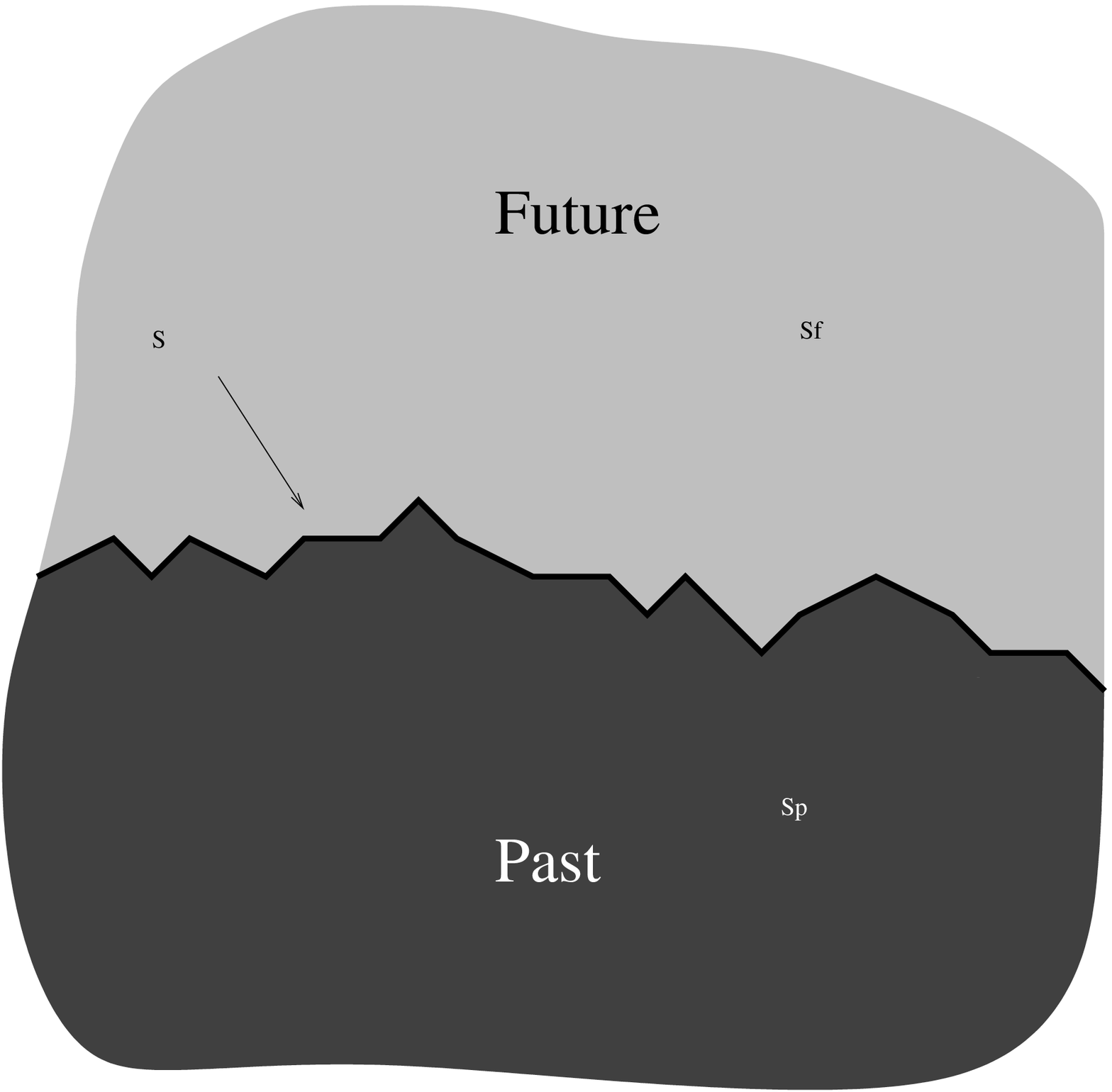} 
\caption{\small Hypersurface $\Sigma_k$ separating `past' and `future' region at step $k$.}\label{hyp}
\end{center}
\end{minipage}
\hspace{.2in}\begin{minipage}{4.5in}\begin{center}      
       \psfrag{sk}{\small$\Sigma_k$}
\psfrag{sk1}{\small$\Sigma_{k+1}$}
\psfrag{t}{}  $
\begin{array}{cccc}
\text{3D perspective:}&\hspace{.5cm}\includegraphics[scale=.1]{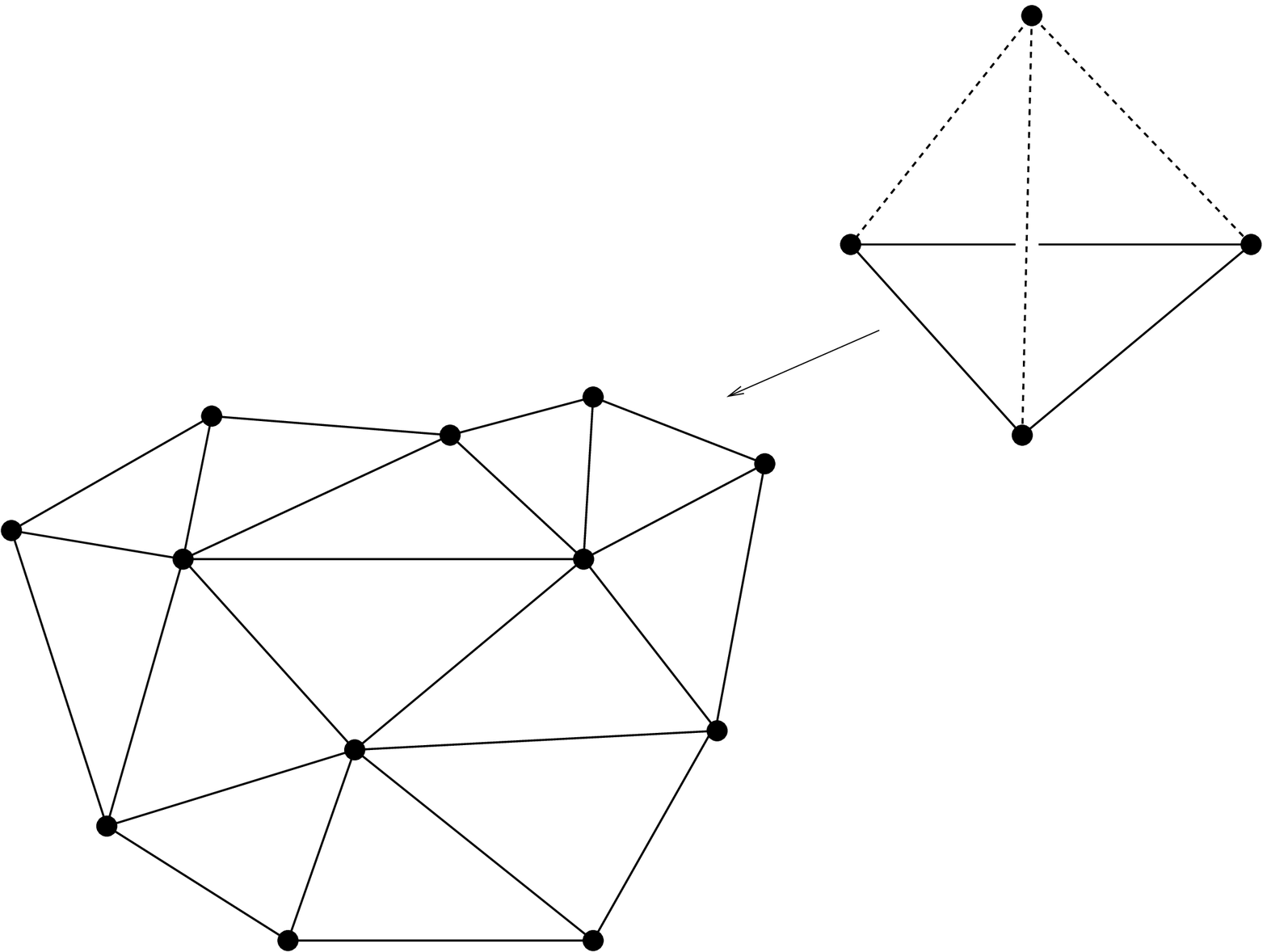}  &\longrightarrow& \hspace{-.7cm}\includegraphics[scale=.1]{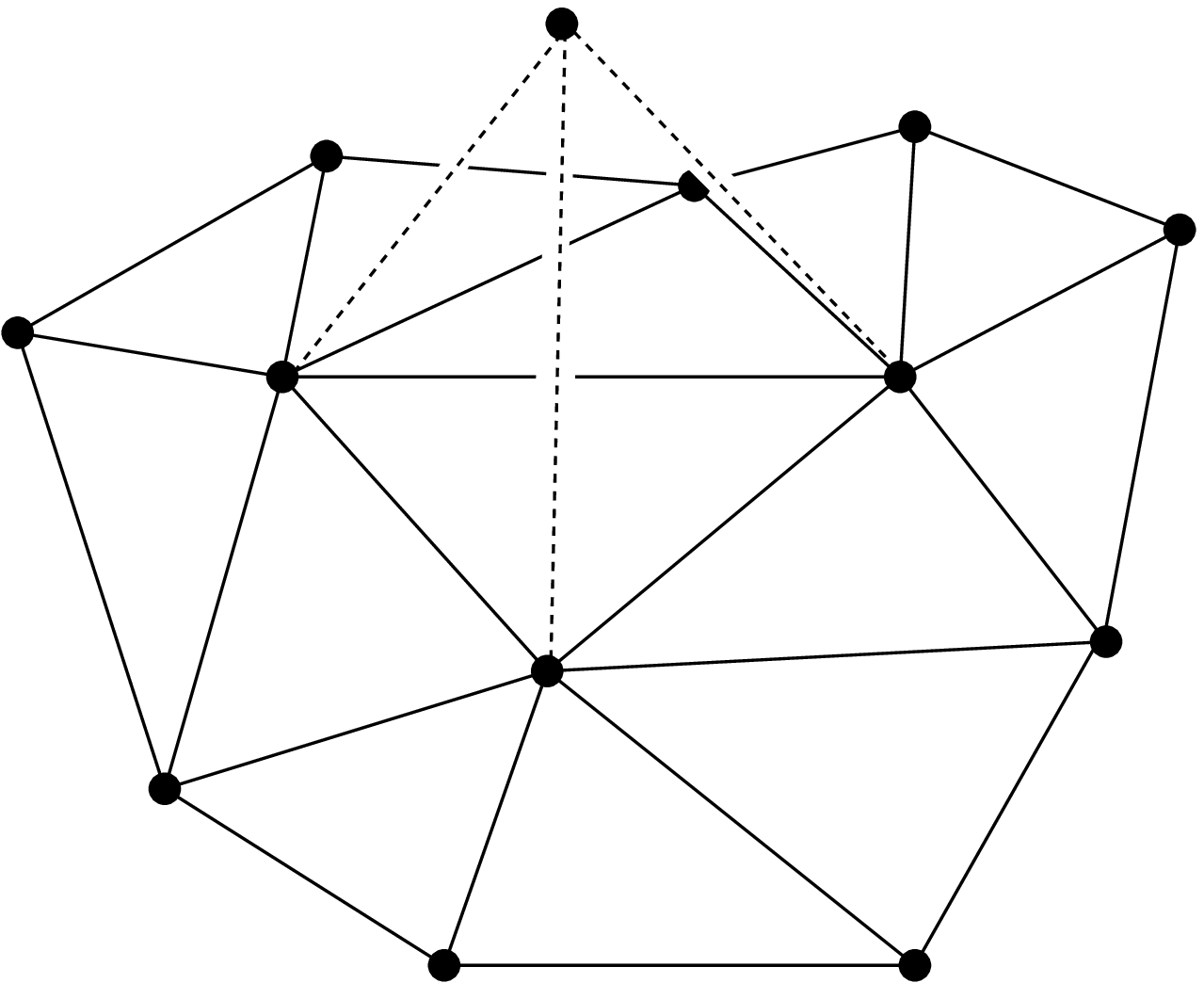} \vspace{.5cm}\\
\text{2D perspective:} &\hspace{.0cm}\includegraphics[scale=.1]{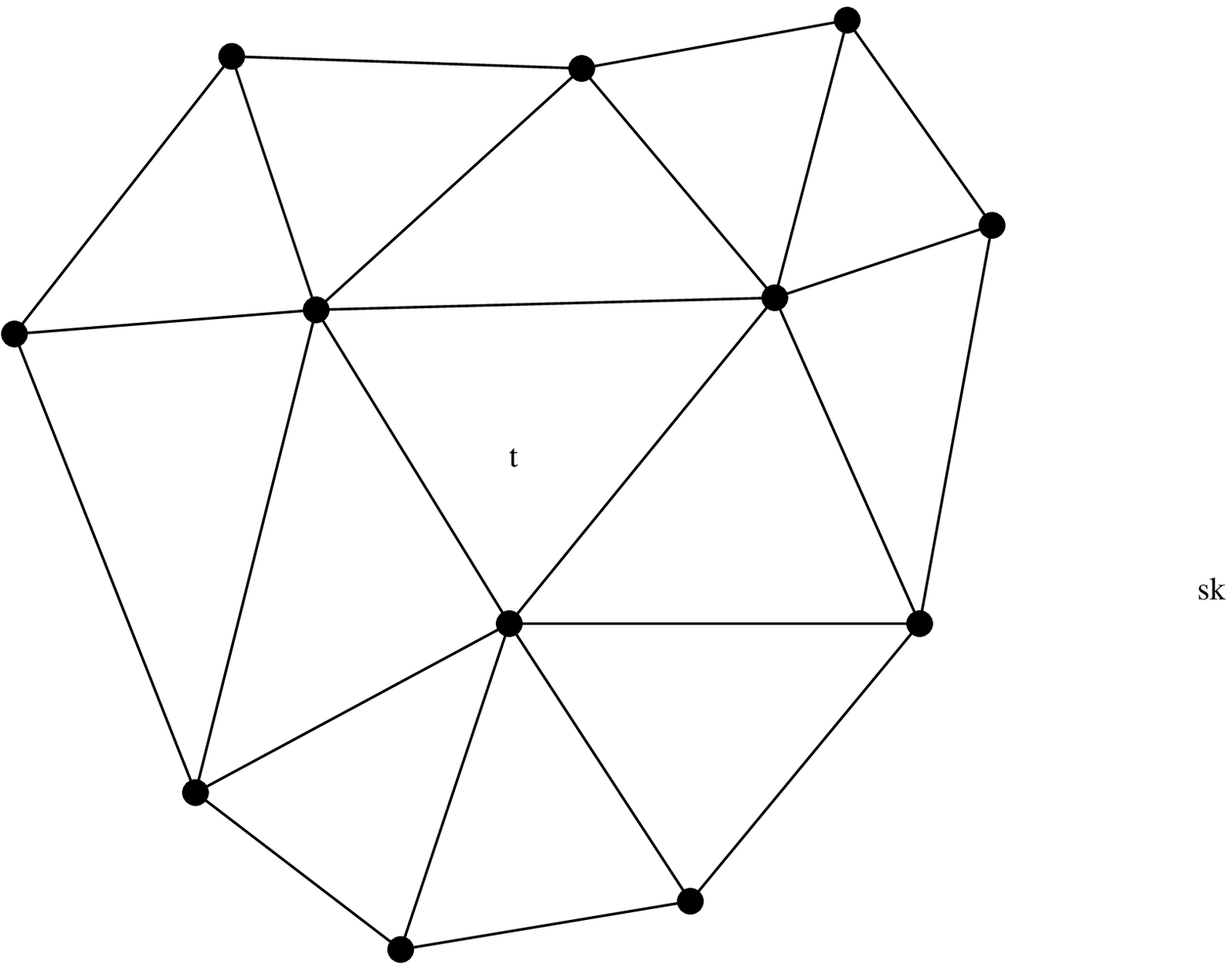} 
&\longrightarrow&\hspace{-.5cm}\includegraphics[scale=.1]{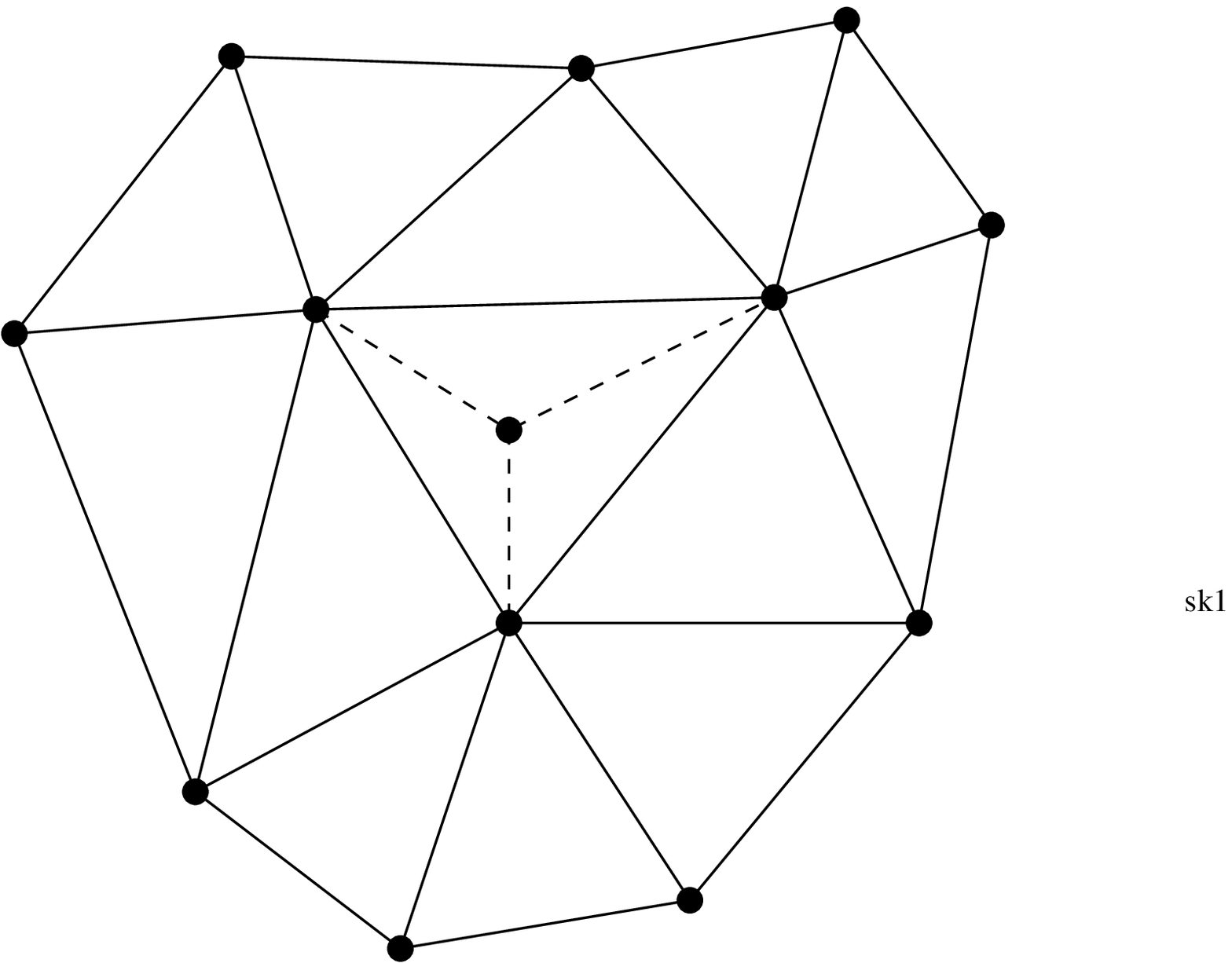}  \\
&& \text{\footnotesize1--3 Pachner move}&
\end{array}$
\caption{\small Gluing a single tetrahedron onto a single triangle in the 2D boundary hypersurface of some 3D bulk triangulation.}\label{pach}
\end{center}
\end{minipage}\end{figure}

The central idea behind the general canonical evolution scheme for triangulations is to glue (or remove) a single $D$--simplex, to (or from) a $(D-1)$--dimensional triangulated hypersurface $\Sigma_k$ at each elementary step counted by $k\in\mathbb{Z}$ (see figure \ref{hyp}); gluing moves correspond to forward evolution, while removal moves correspond to backward evolution. Accordingly, the triangulated hypersurface evolves in a discrete `multi--fingered' or `bubble' time through the full $D$--dimensional solution---in complete analogy to canonical General Relativity. 
The nice feature of these $D$--dimensional gluing or removal moves is that they have the interpretation of $(D-1)$--Pachner moves \cite{pachner} within the $(D-1)$--hypersurface. For instance, figure \ref{pach} shows the 3D example of gluing a single tetrahedron onto a single triangle in a 2D hypersurface. From the 2D perspective, this move amounts to a subdivision of the triangle into three new ones: this is a so--called 1--3 Pachner move in 2D. It turns out that all other allowed gluing and removal moves in 3D and 4D also amount to Pachner moves in the hypersurfaces \cite{csg}. Pachner moves are an elementary class of moves (they only involve a fixed number of simplices) which are ergodic piecewise linear homeomorphisms (one can map between any finite triangulations of the same topology by finite sequences of these moves). These Pachner moves are therefore the general evolution moves which we will henceforth use in our construction of canonical simplicial gravity. 

The implementation of these gluing and removal moves requires the action to be additive. This is, in fact, the case in Regge Calculus on which we therefore put our focus hereafter.\footnote{But note that the present formalism is applicable to any discrete theory with additive action \cite{csg}.} Recall that the lengths of the edges are the configuration variables of Regge Calculus. Thus, when attempting to implement the Pachner moves in a canonical formalism, one, in general, encounters the following two `problems' (which can already be diagnosed for the 1--3 move in figure \ref{pach}):
\begin{itemize}
\item[(a)] subsets of variables coincide at different steps, i.e.\ $\Sigma_{k+1}\cap\Sigma_k\neq\emptyset$, and
\item[(b)] numbers of variables differ (phase space dimension varies) from step to step.
\end{itemize}
The canonical formalism clearly ought to appropriately address and `solve' these `problems'.

\section{Discrete Legendre transformation and implementation of the Pachner moves}

\begin{figure}[htbp!]
\begin{minipage}{2in}
\begin{center}
\psfrag{n}{\tiny $n$}
\psfrag{n+1}{\tiny $n+1$}
\psfrag{sn0}{\tiny $\Sigma_{n-1}$}
\psfrag{sn}{\tiny $\Sigma_n$}
\psfrag{sn1}{\tiny $\Sigma_{n+1}$}
\psfrag{Sn}{\tiny $S_n$}
\psfrag{Sn1}{\tiny $S_{n+1}$}
\includegraphics[scale=.19]{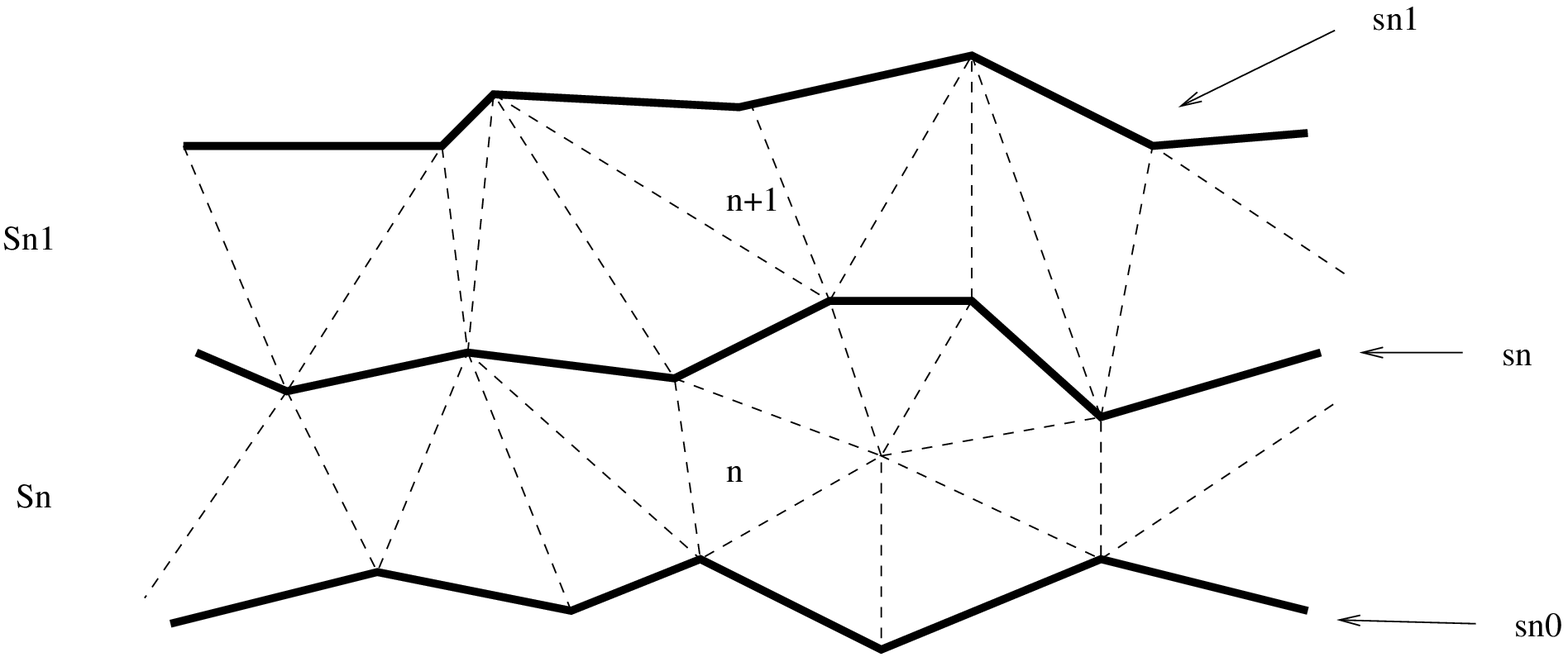}
\caption{\small Group up simplices into a choice of fat slices.}\label{fat}\end{center}
\end{minipage}
\hspace{.1in}\begin{minipage}{2in}
\begin{center}
\psfrag{1}{\tiny 1 - 4}
\psfrag{2}{\tiny 4 - 1}
\includegraphics[scale=.2]{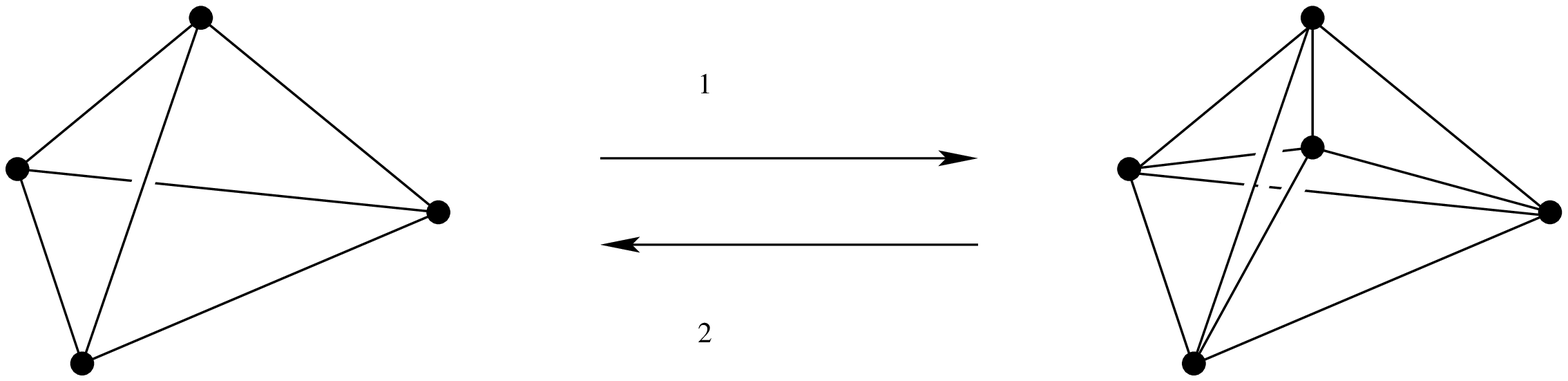}
\caption{\small The 1--4 and 4--1 Pachner moves.}\label{1441}\end{center}\end{minipage}
\hspace{.1in}\begin{minipage}{2in}\begin{center}
\psfrag{1}{\tiny 2 - 3}
\psfrag{2}{\tiny 3 - 2}
\includegraphics[scale=.19]{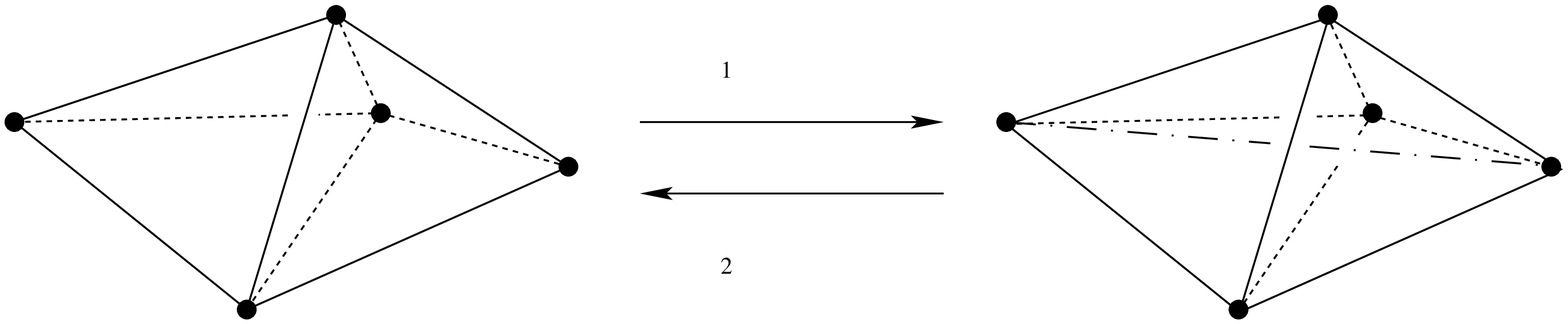}\caption{\small The 2--3 and 3--2 Pachner moves.}\label{2332} \end{center}
\end{minipage}
\end{figure}

In order to define discrete Legendre transformations, we must group up the $D$--simplices counted by $k\in\mathbb{Z}$ into a choice of fat slices which we label by $n\in\mathbb{Z}$ (see figure \ref{fat}). Lengths of edges in $\Sigma_n$ are denoted by $l^e_n$, while all lengths which are internal between $\Sigma_n$ and $\Sigma_{n-1}$ are assigned to step $n$ and denoted by $l^i_n$.\footnote{For simplicity, we neglect any boundary variables.} The key ingredient is to use the action contribution of fat slice $n$, 
$S_n(l^e_n, l^i_n,l^{e'}_{n-1})$, as a `generating function' of first type, defining the conjugate momenta via a pair of discrete Legendre transformations $\mathbb{F}^\pm$ from the direct product of configuration manifolds $\cq_{n-1}\times\cq_n$ to the two phase spaces $\cp_{n-1},\cp_n$ at steps $n-1,n$, respectively, \cite{marsdenwest,csg}
\begin{xalignat}{2}\label{mom}
&{}^+p_e^n\,\,:= \frac{\partial S_{n}}{\partial l^e_n}\,\, \,&&{}^-p_e^{n-1}:= -\frac{\partial S_{n}}{\partial l^e_{n-1}}  \nonumber\\
&{}^+p_i^n\,\,:= \frac{\partial S_n}{\partial l^i_n}  \,\,  \,&&{}^-p_i^{n-1}:=  -\frac{\partial S_n}{\partial l^i_{n-1}}=0  .
\end{xalignat}
The last equation yields a set of constraints which follows from the fact that $S_n$ does not depend on $l^i_{n-1}$. In the general case (e.g.\ due to varying numbers of edges) the Legendre transformations will fail to be isomorphisms and only map onto subspaces of $\cp_{n-1}$ and $\cp_n$ \cite{csg}. We call the image of $\mathbb{F}^-$ the {\it pre--constraint surface} and the image of $\mathbb{F}^+$ the {\it post--constraint surface}. 

From a similar transformation, using $S_{n+1}$ as the generating function, one finds ${}^-p^n_e=-\frac{\partial S_{n+1}}{\partial l^e_n}$. Hence, the equations of motion for the $l^e_n$, i.e.\ $\frac{\partial S_n}{\partial l^e_n}+\frac{\partial S_{n+1}}{\partial l^e_n}=0$, imply a {\it momentum matching} $p^n_e:={}^+p^n_e={}^-p^n_e$. We will therefore subsequently drop the $+,-$ indices. The same applies to internal variables $l^i$, however, now the equations of motion, $\frac{\partial S}{\partial l^i}=0$, imply $p_i=0$, i.e.\ equations of motion for internal lengths actually appear as canonical constraints.

Returning to the gluing moves counted by $k$ (henceforth, we restrict ourselves to gluing moves) and assuming $\Sigma_k=\Sigma_n$, it turns out \cite{csg} that at each $k$ the previous momenta translate into $p^k_e=\frac{\partial S_{k+}}{\partial l^e_k}$, where $S_{k+}=\sum_{k'=1}^kS_{\sigma_{k'}}$ ($S_{\sigma_{k'}}$ action of $k'$-th simplex) is the action up to step $k$ (see figure \ref{hyp}). Since the action is additive, Pachner moves require a {\it momentum updating} \cite{csg}, 
\ba\label{momup}
p^{k+1}_e=p^k_e+\frac{\partial S_{\sigma_{k+1}}}{\partial l^e_k}.
\ea
This construction defines our discrete time evolution. 
 

We are now ready to solve our `problems'. We can solve {`problem' (a)} by mapping the lengths of all edges occurring in both $\Sigma_k$ and $\Sigma_{k+1}$ to themselves, $l^e_{k+1}=l^e_k$, and by updating the momenta according to (\ref{momup}). `Problem' (b) can be solved by a natural phase space extension: we can formally `add' new or old variables $l^n_{k},l^o_k$ of edges occurring only `to the future' or only `to the past' of hypersurface $\Sigma_k$ to $\cp_k$, noting that these variables come with constraints $p^k_n=0=p^k_o$ because of the equations of motion. For the example of the 1--3 Pachner move (figure \ref{pach}), we can use the action of the new tetrahedron, $S_\tau(l^n_{k+1},\ldots)$, as a type one generating function with trivial dependence on $l^n_k$ because the new edges $n=1,2,3$ do not occur at $k$, in order to consistently transform their momenta, $\{p^k_n=\frac{\partial S_\tau}{\partial l^n_{k}}=0\}\mapsto \{p^{k+1}_n=\frac{\partial S_\tau}{\partial l^n_{k+1}}\}$. The first equations are three examples of {\it pre--constraints}, while the last three equations constitute {\it post--constraints} ($S_\tau$ only depends on quantities of $\Sigma_{k+1}$). If this recipe is applied correctly, all Pachner moves in 3D and 4D can be implemented as canonical transformations on (a suitably extended) phase space \cite{csg}.

For 4D simplicial gravity there are four types of Pachner moves \cite{csg}. The 1--4 move (figure \ref{1441}) introduces a new vertex and four new edges which, as in the case of the 1--3 move, come with four {\it post--constraints} $p^{k+1}_n-\frac{\partial S_\sigma}{\partial l^n_{k+1}}=0$ that are automatically satisfied after the move. The 2--3 move (figure \ref{2332}) introduces one edge equipped with a {\it post--constraint}, but renders one triangle internal. In 4D these two moves are the only moves which introduce new edges into the triangulation. However, since these moves do not involve any equations of motion (no edges become internal) all new edge lengths can {\it a priori} be freely chosen. This implies that, with the 2--3 move, one can generate an {\it a priori} free curvature because in 4D Regge Calculus internal triangles carry the deficit angles and the new deficit angle depends on the new length $l^n_{k+1}$. The inverse moves, i.e.\ the 4--1 and 3--2 moves (figures \ref{1441} and \ref{2332}), on the other hand, annihilate edges $o$ from $\Sigma_k$ and come with non--trivial {\it pre--constraints} $p^k_{o}+\frac{\partial S_\sigma}{\partial l^o_k}=0$ which need be satisfied {\it prior} to the move. 

The {\it post--constraints} form an abelian Poisson algebra, but {\it a priori} do not generate gauge transformations \cite{csg}, which in Regge Calculus correspond to vertex displacements in the bulk that leave the action invariant \cite{bd1,dithoe1}. Rather, they reflect the lack of information in a given hypersurface about the full 4D Regge solution; given initial data, non--uniqueness of solutions arises. {\it A posteriori} the first or second class nature of the {\it post--constraints} of $\Sigma_k$ depends on the {\it pre--constraints}: if no complete {\it constraint matching} at $k$ takes place, some constraints turn into second class and the `unmatched' {\it pre--constraints} may fix some (or all) {\it a priori} free lengths of the 1--4 and 2--3 moves \cite{csg,dhta}. Those lengths which eventually remain free constitute gauge parameters. However, gauge symmetry is generically broken in the presence of curvature \cite{bd1,dithoe1}.

\section{Conclusions and Outlook}

We devised a general canonical framework for simplicial gravity by employing gluings/removals of single simplices as evolution moves that can be interpreted as Pachner moves within the hypersurfaces \cite{csg}. The Pachner moves can be implemented as canonical transformations via a natural phase space extension that is controlled by constraints which are equations of motion. By using the action as a generating function, the canonical framework generates general solutions of the covariant formalism and is thus equivalent. Likewise, a quantization should presumably yield a direct connection between the canonical framework and the path integral. The application of the present scheme to linearized 4D Regge gravity, as well as a more detailed discussion and classification of the constraints will appear in a separate paper \cite{dhta}.

{\bf Acknowledgments} The author thanks Bianca Dittrich for collaboration on this topic.

\end{document}